\RequirePackage{graphicx} 
\documentclass[conference]{IEEEtran}
\usepackage{cite}
\usepackage{epsfig}
\usepackage{epstopdf}
\usepackage{graphicx}
\usepackage{url}
\usepackage{amsfonts}
\usepackage{amsmath,bm,amsthm}
\usepackage{amssymb}
\usepackage{multicol}
\usepackage{times}
\usepackage{psfrag}
\usepackage{subfigure}
\usepackage{stfloats}
\usepackage{array}
\usepackage{booktabs, threeparttable}
\usepackage{setspace}
\usepackage{color}
\usepackage[utf8]{inputenc}
\usepackage{bm}  
\usepackage{cuted}
\usepackage{diagbox}
\usepackage{makecell}
\allowdisplaybreaks

\begin{document}
\title{Semi-Supervised Goal-Oriented Semantic Communication Framework for \\Foreground Classification }

\author{
\IEEEauthorblockN{Zhitong Ni\IEEEauthorrefmark{1}, Yansha Deng\IEEEauthorrefmark{2}, and Jinhong Yuan\IEEEauthorrefmark{1} }
\IEEEauthorblockA{\IEEEauthorrefmark{1} School of Electrical Engineering and Telecommunications, University of New South Wales } 
\IEEEauthorblockA{\IEEEauthorrefmark{2}  Department of Engineering, King’s College London    } 
Emails: \IEEEauthorrefmark{1} \{zhitong.ni, j.yuan\}@unsw.edu.au,  \IEEEauthorrefmark{2}  yansha.deng@kcl.ac.uk}

\maketitle

\begin{abstract} 
Wireless goal-oriented semantic communication (GSC) has emerged as a promising paradigm by directly optimizing task performance. However, existing GSC frameworks typically operate on entire images and rely on labeled data for classification tasks, which can limit their compression efficiency and increase the risk of overfitting.
This paper proposes a novel semi-supervised wireless GSC framework for the unlabeled image foreground classification task. In our proposed framework, a foreground-aware masked autoencoder (MAE) is developed to prioritize semantically important foreground objects, thereby reducing transmission overhead.
To enable accurate reconstruction and classification under a limited data size, we further propose a semi-supervised autoencoder (SSAE) that decodes the semantic latent tensor and refines image details by leveraging three complementary information sources, followed by fine-tuning a pre-trained image classification model. The entire pipeline, from foreground masking to classification, is trained in a semi-supervised manner to significantly reduce the need for manual labeling.  
Simulation results validate that the proposed GSC framework achieves over 90\% image classification accuracy while reducing the original image data size by 95\%,  and demonstrate its strong potential for practical tasks in resource-constrained wireless scenarios.
\end{abstract}


\section{Introduction}\label{sec-system}
With the advancement of machine learning technology in wireless communications, semantic communication (SC) has emerged as a pivotal paradigm for next-generation networks \cite{1zhang2022toward}, particularly for bandwidth-intensive applications such as underwater acoustics and satellite communication systems \cite{4chen2025semantic}. Unlike traditional communication systems that focus on the precise transmission of original bits and symbols, SC aims to convey the meaningful source data, thereby offering a compressed data size and enhancing the spectral efficiency of scenarios under stochastic wireless channels with limited resources \cite{5gunduz2022beyond}.

The existing work on SC can be broadly classified into two categories: 
data-oriented \cite{6jiang2022deep} and goal-oriented approaches \cite{deng2026}\cite{3shao2021learning}. The data-oriented SC, such as deep joint source-channel coding  (D-JSCC) \cite{8bourtsoulatze2019deep}, focuses on extracting latent representations from inputs and reconstructing outputs to be as close to the original inputs as possible. Data-oriented SC frameworks typically employ auto-encoders to learn compact latent representations that are transmitted over noisy channels and reconstructed at the receiver with high fidelity  \cite{luoPDFSC}. 
In contrast, goal-oriented SC (GSC) prioritizes the successful execution of a downstream task at the receiver, such as classification or detection, and preserves only the task-relevant semantic representation \cite{10zhang2024unified}. 

Existing GSC frameworks typically compress the entire image (or other media), implicitly assuming that all regions contribute equally to classification performance. However, in practice, different regions of an image contribute unequally, and some may even negatively impact performance \cite{wu2026guaranteed}. This observation motivates foreground extraction to enhance GSC performance.
Another limitation of current GSC frameworks is their reliance on labeled datasets to train the underlying neural networks. In many practical scenarios, however, obtaining comprehensive labels is time-consuming, labor-intensive, and sometimes infeasible.

To address these challenges, we develop a semi-supervised GSC framework for image foreground classification. 
The transmitter is designed to identify and compress foreground objects in the images during the encoding and transmission stages. Then, the receiver obtains the reconstructed images for classifications. 
The entire framework is trained in a semi-supervised manner, where the foreground extraction and reconstruction models are trained in a fully unsupervised manner, while only the final image classification stage requires a limited amount of labeled data. 
 The contributions of this paper are: 

\begin{itemize}
\item At the transmitter, we develop a foreground-aware masked autoencoder (MAE) for semantic information extraction. Unlike conventional masking strategies based on random masking or label-driven masking, our proposed MAE explicitly prioritizes semantically important regions in unlabeled image data. By selectively masking less informative background patches, the MAE is encouraged to learn latent representations that are highly concentrated on foreground content, thereby enabling goal-oriented data transmission.

\item  At the receiver, we propose a semi-supervised auto-encoder (SSAE)  that jointly performs reconstruction and classification of images. Our SSAE reconstructs high-quality foreground images with a given limited data size. Meanwhile, the overall classification performance is improved through the combined contributions of MAE and SSAE.

\item We introduce a semi-supervised training paradigm for the GSC systems. The foreground masking strategy learns semantic representations without requiring annotations; 
the foreground information is compressed and reconstructed in a label-free setting; 
and a pre-trained image classification model is fine-tuned using a limited amount of labeled data.  This training paradigm significantly reduces the reliance on extensive manual labeling and facilitates the practical GSC deployment in real-world scenarios.
\end{itemize}

Notations:   $\rm\bf a$ denotes a vector,
$\rm\bf A$ denotes a matrix or tensor, $[{\bf A}]_{i,j}$ denotes the $(i,j)$-th entry of $\bf A$. Italic letters like $N$ and  $\alpha$ are scalars.  $\|{\rm\bf A }\| $ is the Frobenius norm of a matrix or a tensor. Operator   $ \circ $  denotes the Hadamard product. $ {\mathbb E}(\cdot)$  obtains the expectation of a random scalar.

\section{System Model}\label{SyS}
 
\begin{figure*}[tb]
	\centering 
	\includegraphics[width=1\linewidth]{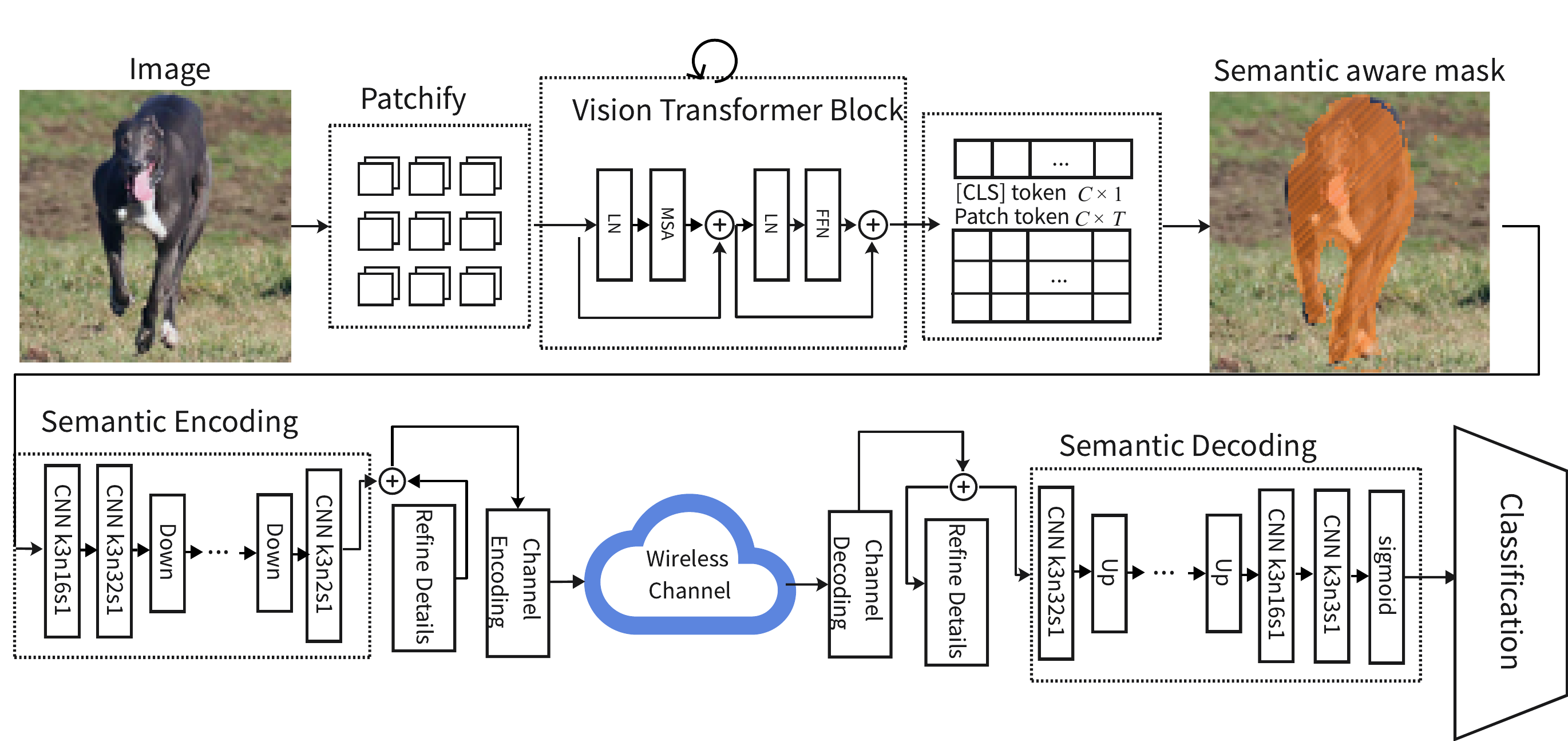} 
	\caption{Illustration of our proposed image classification framework, where the image first goes through a ViT-based model to detect the semantically meaningful foreground, then goes through a semantic encoder for the image's foreground reconstruction, and the encoded outputs are transmitted in the wireless channel. At the receiver, the image foreground is extracted using a symmetric model and then passed through a classification model in a semi-supervised manner.}
	\label{sysmodel}
\end{figure*} 
In this section, we introduce the architecture of our proposed GSC system. The main goal of this system is to reconstruct the image of interest and classify the images within limited categories.
 
The information source is an image tensor ${\bf I}$ of dimension  $3\times H\times W$, where the $3$ channels correspond to red/green/blue (RGB) color channels, and $H$ and $W$ represent the height and width of the images, respectively. 
 As shown in Fig. \ref{sysmodel}, the transmitter extracts the image foreground. Several modules sequentially process the foreground to produce the semantically encoded latent representation, ${\bf S}$. This latent variable is then channel-encoded and transmitted over the wireless channel. At the receiver, the recovered latent variable is decoded for image reconstruction and classification.

Firstly, we aim to obtain the foreground using the vision Transformer (ViT)-based MAE, of which details will be given in Section \ref{SMAE}.
Secondly, after obtaining the semantic mask, denoted as ${\bf M}$, the image tensor is masked to maintain the foreground only, and goes through the semantic encoding, i.e., SSAE,  to further reduce the data size while maintaining high fidelity.
Finally, the data stream to be transmitted after the channel encoding is 
\begin{align} 
{\bm x} ={\mathcal C}( {\bf S} ) 
 ={\mathcal C}({\mathcal E} ({\bf M}^{(3)}\circ{\bf I})),
\end{align}
where $\mathcal C(\cdot)$ denotes the conventional channel encoding operation, ${\mathcal E}(\cdot)$ is the SSAE  as will be detailed in  Section \ref{SSAE}, and ${\bf M}^{(3)}$ is a 3-dimensional (3D) tensor with each channel being ${\bf M}$.
 
For wireless communications, we use the additive white Gaussian noise (AWGN) channel, which is widely adopted for justifying the SC systems. The received data symbols can be given by
\begin{align} 
{\bm y}  =   {\bm x} + {\bm n}, 
\end{align}
where  ${\bm n}$ is the AWGN of zero mean and variance of $\sigma^2$.

At the receiver, the received signal is processed by the channel decoder to correct bit errors, followed by the semantic decoder to reconstruct the transmitted foreground for image classifications.  
The reconstructed image is given by
\begin{align} 
  \hat {\bf I}={\mathcal E}^{-1}({\mathcal C}^{-1}({\bm y})),
\end{align}  
where $ {\mathcal C}^{-1} (\cdot)$ is the channel decoder and ${\mathcal E}^{-1}(\cdot)$ is the semantic decoder.  
An image classification model is trained to identify $C'$ classes, indexed by $c = \{1,  \cdots, C'\}$. Given the reconstructed image, the classification model outputs a probability distribution ${\bf p}=\{{p}_1,\cdots, {p}_{C'}\}$, where ${p}_c$ represents the probability of recognizing $\hat {\bf I}$ as class $c$.

\section{Semi-Supervised GSC Framework }\label{MAE} 
In this section, we illustrate the framework of semi-supervised GSC that extracts the foreground of the image, obtains the compressed image latent representation, and classifies the reconstructed image. 

\subsection{MAE}\label{SMAE}
We aim to obtain a semantic mask that removes the background of the image since the background generally provides no or negative contributions to classifying the specific class.  

Due to the unavailability of image labels, we patchify the image and leverage the inherent and coherent semantic information of each separated patch of the image.
Each patch has the size of $3\times P \times P$, where $P$ is the patch size. Note that $H$ and $W$ should be integer multiples of $P$, and the number of patches is $T =HW/ P^2$. 

Each image patch can be transformed into a meaningful token that describes partial information about the image. 
To transform the image patches into tokens, we adopt the ViT architecture as in \cite{caron2021emerging}, where the images are sequentially processed by a convolutional neural network (CNN) and multiple Transformer blocks. It is well known that the Transformer can capture long-range dependencies between image patches through its self-attention mechanism, enabling more effective feature representations.

The output of the final Transformer block is a $(T+1)\times C$ token matrix $\bf T$ that consists of one classification token, commonly referred to as the ``[CLS]'' token, and $T$ patch tokens that correspond to each image patch.  
The dimension of the token is $C\times 1$, which determines the semantic representation capability of the token.
Since position embedding is adopted for the patch tokens to represent their specific image region, it introduces spatial bias into patch tokens, which can interfere with learning position-invariant semantic representations that are essential for foreground extraction.
The [CLS] token that has no position embedding aggregates global semantic information from the entire image. 
This unique token can serve as a robust semantic query about the foreground object. 

Now, suppose that the [CLS] token represents the global semantic information regarding the entire image by using a pre-trained ViT. The problem is to remove the background information from the [CLS] token. 
We project the [CLS] token produced by the final Transformer block into a high-dimensional embedding space. This is achieved by appending a multi-layer perceptron (MLP) projection head to the [CLS] token. The embedding dimension is thus increased from $C$ to $K$. 
The output of the MLP is normalized by a temperature-scaled softmax function to make the output become a probability distribution function (pdf), i.e.,
\begin{align} 
   q_k = \frac{e^{Q_k/\epsilon}}{\sum\limits_{k=1}^{K} e^{Q_k/\epsilon}},
\end{align}
where $Q_k$, $Q_k\geq 0$, is the output of the MLP and $\epsilon$ is a temperature scaler.  The entire neural network mapping from the input image ${\bf I}$ to the softmax-normalized output $q_k$ forms the entire neural network architecture of the MAE, denoted as ${\mathcal N}$.

The pdf $q_k$ is determined by the weights of ${\mathcal N}$ and the input ${\bf I}$. To generate a foreground-aware output in a supervised manner, we need to process the image into a perfectly masked input, i.e., ${\bf I}\circ {\bf M}^{(3)}$, and obtain the corresponding output as labels, and then train the network ${\mathcal N}$. However, this is not a practical solution since a perfectly masked input needs extremely intense labor.

This motivates us to generate a semantic mask in a self-supervised learning manner that is inspired by the distillation with no label (DINO) framework \cite{caron2021emerging}.  Specifically, two different views of the same image ${\bf I}$ are constructed, denoted as ${\bf I}_s$ and ${\bf I}_t$, where ${\bf I}_t$ is a pre-masked image and ${\bf I}_s$ conducts additional data augmentation on ${\bf I}_t$. 
In our paper, ${\bf I}_t$ is obtained using a pretrained DINO teacher network, ${\mathcal N}_t(\cdot)$, to generate a coarse image mask on ${\bf I}$, i.e., ${\bf I}_t={\mathcal N}_t({\bf I})$. ${\bf I}_s$ is obtained by conducting additional random cropping and color jittering on ${\bf I}_t$. The cropping is realized by randomly masking $X$ image patches on ${\bf I}_t$. The color jittering is done by adjusting the brightness by $\xi$ times on ${\bf I}_t$.
${\bf I}_s$ is processed by a student network ${\mathcal N}_s(\cdot)$ that will be trained.  

The training objective is to minimize the cross-entropy loss between the outputs of  ${\mathcal N}_t({\bf I}_t)$ and ${\mathcal N}_s({\bf I}_s)$. Hence, the loss function of the MAE is   
\begin{align}
    {\mathcal L} = - \sum\limits_{k=1}^{K} q_{k,t}\log  q_{k,s},
\end{align}
where $q_{k,t}$ and  $q_{k,s}$  are the outputs of ${\mathcal N}_t({\bf I}_t)$  and ${\mathcal N}_s({\bf I}_s)$, respectively.
The gradients are propagated only through the student network, while ${\mathcal N}_t$ is kept fixed during backpropagation.  

After training, the student network is expected to produce outputs similar to those of the teacher network, since the cross-entropy loss is minimized when $q_{k,s} = q_{k,t}$ for a given $q_{k,t}$. However, the student network has access to less information than the teacher network. Hence, the output of the student network can have more focus on the foreground-relevant features by leveraging the Transformer's attention mechanism.
When a fully detailed image ${\bf I}$ is fed into the student network, the produced [CLS] token therefore contains less redundant background information, whereas the original DINO network has both foreground and background information.

Recall that each patch token corresponds to a specific spatial region. By leveraging both the [CLS] token and the patch tokens, we can generate the final semantic mask ${\bf M}$ through the self-attention weights, as detailed in Appendix \ref{App1}.
 
\subsection{SSAE}\label{SSAE}
 
This subsection develops a novel SSAE that reconstructs the image from limited data bits.
To enable reliable image classification at the receiver, conventional image compression methods such as JPEG can be employed to reconstruct the image under a limited data size. However, the JPEG format can be corrupted under low signal-to-noise ratio (SNR) due to its variable-length Huffman coding.

Given the strong capability of CNNs in modeling spatial correlations and reconstructing high-dimensional signals, we are inspired by the architecture in \cite{wang2019} and  \cite{8bourtsoulatze2019deep}, and propose an encoder setup as shown in the ``Semantic Encoding'' in Fig. \ref{sysmodel}, where in each ``CNNk3n$x$s1'' module, `k', `n', and `s' denotes the kernel size, the number of output channels, and the padding size, respectively, and the stride is 1. 
Each of these CNN modules, except the final CNN module in the penultimate layer of the decoder, connects a batch normalization layer and a rectified linear unit (ReLU), which are omitted in Fig. \ref{sysmodel}. The ``Down'' modules consist of a pooling layer followed by a ``CNNk3n32s1'' module, where the pooling layer has a convolutional kernel of size $2$, a stride of $2$, no padding, and a group of $32$, which halves the spatial dimensions and maintains the channel dimension.  
The decoder is at the receiver side and has symmetric setups.  After the final CNN module of the decoder, a sigmoid function is adopted to constrain the values between 0 and 1. 

The input to the SSAE should be the masked region of the image only, which does not form a rectangular shape. To further reduce the transmission cost, we additionally mask $T'$ patches that have the largest distance from the foreground centroid. These $T'$ patches will be refined using the methods at the end of this subsection. For the rest of the foreground, the SSAE in Fig. \ref{sysmodel} progressively transforms the masked image into a low-dimensional latent tensor at the bottleneck, which serves as a compressed latent representation of the image foreground. The total dimension of the CNN layers is dropped from $3HW$ to $C_o\frac{H'}{2^D}\frac{W'}{2^D}$, where $H'$ and $W'$ are the masked foreground's spatial dimensions, $C_o$ is the bottleneck's channel dimension, and  $D$ is the number of Down modules. 

Now, we can generate a semantically rich and quantized latent tensor at the bottleneck. For wireless communications, the continuous-valued latent representation must be quantized before transmission.  Hence, we apply $N$-bit element-wise quantization in the entire range of the latent values.

At the receiver side, we send the quantized latent tensor to the decoder.  
For training the entire CNN model, the loss function is the mean squared error of the masked foreground. It is noted that we omit the wireless channel and noise, and connect the encoder and decoder directly for offline training.  This is because channel coding is used to mitigate the noise influences. We still need to consider the quantization of the latent tensor during training. 
In backpropagation, we approximate the gradient of the quantization function by a straight-through estimator (STE), which is widely adopted for training models with discrete or quantized latent representations.

The CNN-based image reconstruction achieves performance gains under a limited data size and low SNRs. Recall that there are $T'$ patches that need to be refined, since they are away from the image centroid and make the latent values become redundant. For reconstructing all the non-masked patches, including both CNN-decoded and refined ones, we need to transmit a binary sequence of length $T$, where 0 indicates the patch is masked and 1 indicates the patch is non-masked. At the receiver side, 
We define the refinement ratio as  
\begin{align}
\eta =  {T'}/M,
\end{align} 
where $ M$ is the number of non-masked patches. The refinement ratio measures the ratio between the CNN-decoded patches and refined ones. After receiving the $T$ bits, the receiver can obtain the positions of both CNN-decoded and refined patches.
When $\eta$ is 0, the receiver fully uses the CNN's latent representations to reconstruct the image. Otherwise, for those $T'$ patches, we use the traditional K-means method to refine them.

The K-means method can obtain a palette of $F$ representative cluster centers among $T'$ patches, where each cluster center corresponds to a 3-byte representative color. Each pixel in the refined patch is then replaced by the index of its nearest cluster center (an integer in $\{0,\cdots, F-1\}$), effectively quantizing the patch using the learned palette. The indices of the refined patches are further compressed via run-length encoding (RLE) of run length $L$, leveraging the spatial redundancy in palette-based representations.

In summary, the transmitter thus needs to send three key components to the receiver:
\begin{itemize}
\item The quantized latent tensor of the non-masked images obtained by the CNN encoder. 
\item The short binary sequence of length $T$ indicating which patches need refinement.
\item The color index for the refined patches followed by the palette.
\end{itemize}

Finally, we fine-tune a pre-trained mature classification model to classify the reconstructed images. The adopted classification model has a similar architecture to the MAE used in our framework.  In the classification model, the [CLS] token is projected onto the $C'$ categories instead of the original $K$-dimensional embedding space, followed by using the cross-entropy loss function. Since fine-tuning requires labeled data, we use a small amount of labeled data to train the classification model. Consequently, for the entire pipeline from masking to classification, only a little human labeling is required at the final stage of image classification. This significantly reduces manual labeling effort, making our semi-supervised learning approach avoid extensive human intervention.

\section{Simulations} 
 
In this section, we present numerical results to illustrate the performance of our semi-supervised GSC framework.  
The experiments are conducted on the STL-10 dataset, a widely used benchmark for image reconstruction and representation learning tasks. 
The dataset consists of 100,000 unlabeled natural images collected from 10 object categories, 5,000 labeled images for supervised learning, and 8,000 labeled images for testing. This unlabeled split is adopted for our self-supervised learning of foreground extraction and image reconstruction, and the labeled images are used for fine-tuning the image classification models. 
 
\begin{figure}[tb]
	\centering 
	\includegraphics[width=0.8\linewidth]{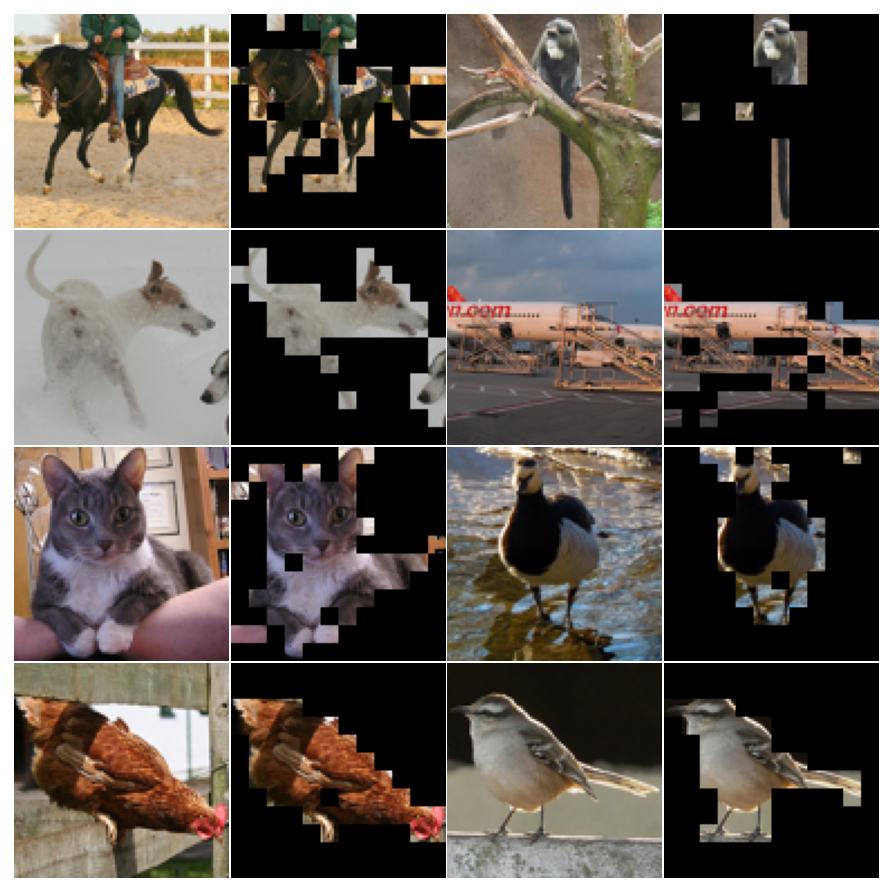} 
	\caption{Illustration of the MAE, where two pictures are a comparison pair, with the left one and right one being the original picture and foreground, respectively.}
	\label{fig1}
\end{figure}  
Fig. \ref{fig1} shows representative image pairs illustrating the foreground extraction capability of our proposed MAE. The image spatial dimension is $96\times 96$, the patch size is $P=8$, and thus, the image is divided into $T=144$ patches. 
The number of multi-head attentions is $5$ for ${\mathcal N}_s$, and we select 5 out of 6 attention heads except the 2nd one in the DINO pretrained network ${\mathcal N}_t$. We set $X=10$ and $\xi\in[0.9,1.1]$ for training our MAE. The threshold for truncating the mask $\rho$ is 0.004.  Other parameters are the same as in the DINO network. 
The results are illustrated in pairs with the original input image (left) and its corresponding masked output (right). The masking process preserves both fine- and large-scale semantic components, such as animal contours and main bodies. These qualitative examples demonstrate that our proposed MAE effectively captures high-level contextual information while maintaining visually meaningful content across diverse object categories, scales, and scene complexities.
We further measure the proportion of non-masked pixels in the entire dataset. Approximately 49\% of the pixels remain non-masked, indicating that nearly half of the image can be removed, which is beneficial for subsequent compression and transmission efficiency.

\begin{figure}[tb]
	\centering 
	\includegraphics[width=1\linewidth]{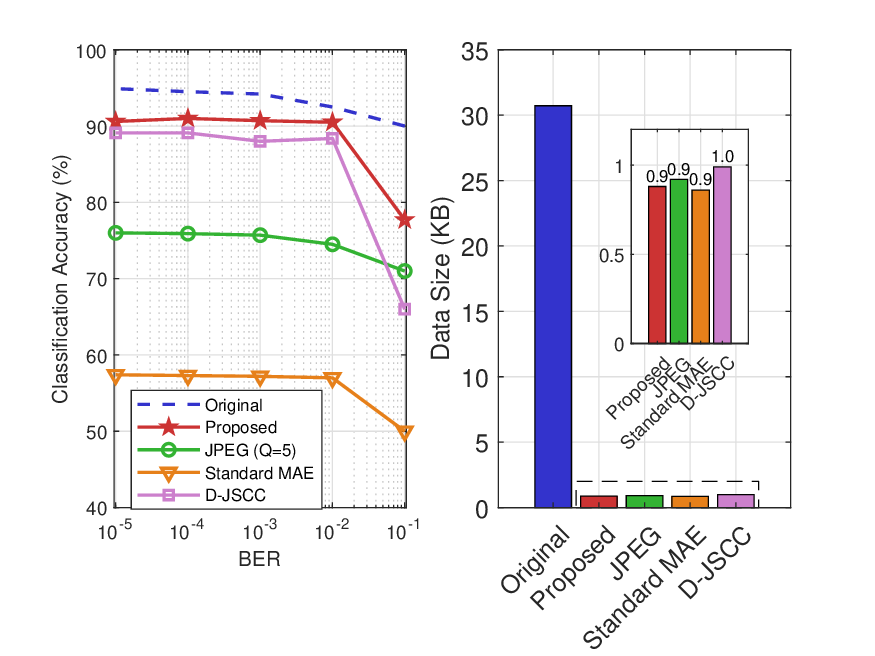} 
	\caption{Classification accuracy of different image reconstruction models. The data size refers to the number of bit streams before channel encoding. }
	\label{fig3}
\end{figure}

Fig. \ref{fig3} illustrates the classification accuracy of the reconstructed images versus the bit error rate (BER), where the BER refers to the bit error rate after the image is reconstructed.  
We compare our methods with the JPEG format with a quality of 5 (Q=5), the standard MAE with JPEG compression \cite{he2022masked}, and the D-JSCC \cite{8bourtsoulatze2019deep}. 
We adopt a pre-trained ViT classification model originally trained on ImageNet with 1000 classes, and modify its classification head to accommodate our 10 object categories. The training data is based on the STL-10 training dataset. 
The data size is constrained to around 0.9-1.0 KB. In our proposed MAE+SSAE, we adjust the model parameters to achieve the highest classification accuracy under the constrained data size, which are shown as follows: We let $C_o=2$,  $D=2$, and $N=7$,  and set $\eta \approx 0.4$, $F=8$, and $L=4$, respectively. 
As shown in the figure, the classification model is robust to the BER below $10^{-2}$. Our proposed method achieves significantly higher accuracy than the baselines under similar data-size constraints. 
The original STL-10 test images achieve the highest accuracy of around 94\% when the BER is below $10^{-3}$, with accuracy degrading as the BER increases. However, the original images require $96\times 96\times 3  /1024 \approx 27$ KB. In contrast, our proposed framework reduces the data size to 0.9 KB, which saves around 95\% of transmission costs, while the classification accuracy is only 4\%-5\% lower than that of the original one.

\begin{table}[t] 
\begin{center}
\caption{Performance of Image Reconstructions Under AWGN}\label{tab1} 
\begin{tabular}{c|ccccccc}
\hline
  \diagbox{Models}{PSNR(dB)}{SNR}   &    \makecell{ 15 dB  }  &    \makecell{ 10 dB  }  & \makecell{ 5 dB } \\ 
\hline    
 Proposed  &    25.4 &  23.9  & 12.8  \\  
Standard MAE \cite{he2022masked} &    16.0   & NaN &  NaN   \\
D-JSCC \cite{8bourtsoulatze2019deep} &   23.7  &  22.5 & 12.0    \\
 JPEG (Q=5)  &  22.0 & NaN &  NaN \\
 Original    &  $\infty$  & 25.8  &  19.2   \\
\hline
\end{tabular}
\end{center} 
\end{table}

Table \ref{tab1} illustrates the performance of image reconstructions against AWGN.  
The performance metric is the peak-signal-to-noise ratio (PSNR), which is defined as
\begin{align}
10 \log_{10} \left( 
\sum\limits_{i,j,c=1,1,1}^{H,W,3}\frac 1{3 P^2 M}
\left( I_{i,j,c} - \hat{I}_{i,j,c} \right)^2   M_{i,j} 
\right)^{-1},
\end{align}
where $I_{i,j,c}$ is the $(i,j)$-th original pixel in the $c$-th color channel, $\hat I_{i,j,c}$ is the corresponding estimated pixel,  and $M_{i,j}$ is the masking coefficient. The SNR is given by ${\mathbb E}(|\bm x|^2)/\sigma^2$. For channel encoding, direct bit streams with binary phase shift keying are adopted. 
From the table, we see that the original image has no reconstruction error at an SNR above 15 dB since the bits are correctly received. When SNR drops to 10 dB, JPEG-based bit streams are highly fragile to noise, and some figures cannot be reconstructed due to variable-length Huffman coding. This often results in complete decoding failure or severe image corruption.  
For the D-JSCC, its performance remains lower than that of our proposed MAE+SSAE across the considered SNR range.  
Overall, our proposed method achieves the highest reconstruction quality among the baseline methods and demonstrates strong robustness to noise.

\section{Conclusions} 
In this paper, we proposed a novel GSC framework for image transmission and classification over wireless channels, incorporating foreground extraction, compression, reconstruction, and classification. By prioritizing semantically important foreground regions using a ViT-based MAE, our proposed framework effectively removes background information.
Our proposed SSAE enables highly compact yet semantically informative latent representations for unlabeled image datasets and outperforms the state-of-the-art methods in terms of classification accuracy and PSNR.
In summary, our proposed approach offers a practical and efficient solution for joint image reconstruction and classification in bandwidth-constrained scenarios.

\begin{appendices}
\section{Generations of Attention Maps}\label{App1}
For a token block ${\bf T}$ of size $  (T+1) \times C$, we obtain the multi-head self-attention (MSA) matrix as 
\begin{align}
 {\rm MSA}({\rm LN}({\bf T}  ))  ,
\end{align}
where  ${\rm LN}(\cdot)$ is the layer normalization of $ {\bf T}$, such that each row of ${\bf T}$ is normalized to have zero mean and variance of 1.
 
The MSA  operation involves linear projections, scaled dot-product attention calculations, and concatenation. For the MSA mechanism, the input ${\bf T}$  is transformed into Query and Key subblocks.
For each attention head $m$, $m\in\{1,\cdots, M'\}$, the Query (Q) and Key (K) subblocks are
\begin{align}
 {\bf Q}_m= {\bf T}  {\bf W}_m^Q,  {\bf K}_m= {\bf T}  {\bf W}_m^K,   
\end{align}
where ${\bf W}_m^Q$ and ${\bf W}_m^K$ are $C\times  C/M'$ matrices whose weights are given by the last Transform block in Fig. \ref{sysmodel}. Then, the $m$th MSA matrix is 
\begin{align} \label{AttMtx}
{\bf S}_m = {\rm softmax}\left( {{\bf Q}_m  {\bf K}_m^T}  \right) / {\sqrt{C/M'}}, 
\end{align}
where ${\rm softmax}(\cdot)$ is the softmax function. Then, we obtain the first row that corresponds to the attention from the [CLS] token to each patch token, as shown as
\begin{align} \label{AttMtx}
{\bf a}_m = [{\bf S}_m]_{1,2:(T+1)}.
\end{align}
The attention map is a $H/P\times W/P$ matrix that is reshaped from the $1\times T $ vector ${\bf a}_m$,  denoted as ${\bf A}_m$.
By concatenating all $M'$ MSA heads,  the multi-head attention map is ${\bf A}={\rm cat}({\bf A}_1,\cdots,{\bf A}_{M'})$ that has the dimension of $M'\times H/P\times W/P$. 

To obtain the pixel-wise semantic mask ${\bf M}$, we directly sum the multi-head attention maps across the head dimension, followed by upsampling to the original image size $H\times W$ using a simple bilinear interpolation operation.
We set the entries larger than $ \rho $ to be 1s and all others to be 0s.

\end{appendices}
\bibliographystyle{IEEEtran} 
\bibliography{ref}
\end{document}